\documentclass[12pt]{article}
\usepackage{graphicx}
\input epsf
\usepackage{amssymb}
\parskip=\medskipamount
\def\om{\omega}   
      
\def\uno{\relax{\rm 1\kern-.18 em l}}

\def\IK{\relax{\rm l\kern-.18 em K}}
\def\IL{\relax{\rm I\kern-.18 em L}}

\def\ii{\rm i\,}

\def\dfrac#1#2{{\displaystyle\frac{#1}{#2}}}

\def\frac#1#2{{#1\over #2}}

\def\ptos{\leaders\hbox to 2mm{\hfil{.}\hfil}\hfill}
\def\\{\hfill\break}

\def\<#1>{\langle#1\rangle}
\def\ii{{\rm i\,}}
\font\tenfrak=eufm10  \font\sevenfrak=eufm7  \font\fivefrak=eufm5
\newfam\frakfam
\textfont\frakfam=\tenfrak\scriptfont\frakfam=\sevenfrak
\scriptscriptfont\frakfam=\fivefrak

\font\tengoth=eufm10 scaled\magstep1 \font\sevengoth=eufm7
\font\fivegoth=eufm5
\newfam\gothfam
\textfont\gothfam=\tengoth\scriptfont\gothfam=\sevengoth
    \scriptscriptfont\gothfam=\fivegoth

\newtheorem{proposicion}{Proposition}

\def\today{\ifcase\month\or
     January\or February\or March\or April\or May\or June\or
     July\or August\or September\or October\or November\or December\fi
     \space\number\day, \number\year}

\begin{document}

\title{A quantum exactly solvable non-linear oscillator related
with the isotonic oscillator  }

\author{J.F. Cari\~nena$\dagger\,^{a)}$, A.M. Perelomov$\ddagger\,^{b)}$,
M.F. Ra\~nada$\dagger\,^{c)}$, \\ and M. Santander${\S}\,^{d)}$ \\
$\dagger$
   {\sl Departamento de F\'{\i}sica Te\'orica and IUMA, Facultad de Ciencias} \\
   {\sl Universidad de Zaragoza, 50009 Zaragoza, Spain}  \\
$\ddagger$
   {\sl Departamento de F\'{\i}sica Te\'orica, Facultad de Ciencias} \\
   {\sl Universidad de Zaragoza, 50009 Zaragoza, Spain}  \\
{\S}
   {\sl Departamento de F\'{\i}sica Te\'orica, Facultad de Ciencias} \\
   {\sl Universidad de Valladolid,  47011 Valladolid, Spain}
}
\date{}

\maketitle

\begin{abstract}
A nonpolynomial one-dimensional quantum potential representing an
oscillator, that can be considered as placed in the middle between
the harmonic oscillator and the isotonic oscillator (harmonic
oscillator with a centripetal barrier), is studied. First the
general case, that depends of a parameter $a$, is considered and
then a particular case is studied with great detail. It is proven
that it is Schr\"odinger solvable and then the wave functions
$\Psi_n$ and the energies $E_n$ of the bound states are explicitly
obtained. Finally it is proven that the solutions determine a
family of orthogonal polynomials ${\cal P}_n(x)$ related with the
Hermite polynomials and such that: (i) Every ${\cal P}_n$ is a
linear combination of three Hermite polynomials, and (ii) They are
orthogonal with respect to a new measure obtained by modifying the
classic Hermite measure.
\end{abstract}

\begin{quote}
Nonpolynomial potentials.

{\sl Running title:}{\enskip}
A quantum exactly solvable non-linear oscillator.


Dirac, Klein-Gordon
\end{quote}

{\vfill} \footnoterule {\small $^{b)}$ {On leave of absence from
Institute for Theoretical and Experimental Physics, 117259 Moscow, Russia} \\
{\it E-mail :} $^{a)}$ {\tt jfc@unizar.es} ;
$^{b)}$ {\tt perelomo@dftuz.unizar.es} ; $^{c)}$ {\tt mfran@unizar.es} ; \\
$^{d)}$ {\tt msn@fta.uva.es}.
\newpage

\section{Introduction}

  It is well known that, in quantum mechanics,  the family of
Schr\"odinger solvable potentials is very restricted and also that
exact solvability is a very delicate property (see Ref.
\cite{OlPe83} for a review and Refs.
\cite{DuGaRS01}--\cite{CrNeNi07} for some recent papers on this
matter). In fact, in most of cases, the addition of a small
perturbation to a quantum solvable system breaks this property and
leads to potentials that must be analyzed by the use of
perturbative methods, variational formalisms or numerical
techniques. The most interesting and best known system inside this
small family is the harmonic oscillator whose energy spectrum
consists of an infinite set of equidistant energy levels. Many
other oscillators, as for example harmonic oscillators perturbed
by a term containing a fourth or a sixth power in the coordinate,
have been extensively studied but making use of the above
mentioned techniques. Nevertheless, it is known the existence of
another oscillator, the so-called isotonic oscillator, that is
exactly solvable and is endowed with many interesting properties.

  The main aim of this paper is to present an study of a new
nonpolynomial one-dimensional quantum potential representing an
oscillator, that can be considered as placed in the middle between
the harmonic oscillator and the isotonic oscillator. We will prove
that is exactly solvable and that the energy spectrum and the wave
functions have properties closely related with those
characterising the harmonic oscillator.

  In more detail, the plan of the article is as follows: In Sec. II
we recall the main characteristics of the isotonic oscillator  In
Sec. III we present a new potential depending of a parameter $a$;
first we study the general case and then we solve the particular
case $a^2=1/2$ obtaining the wave functions and energy spectrum.
Sec. IV is devoted to analyze a family of orthogonal polynomials
and to study the relation with the Hermite polynomials. Finally,
in Sec. V we make some final comments.

\section{The Isotonic oscillator}
The following potential
$$
  U_{\rm Isot}(x) = U_0(x) + U_g(x) = (\frac{1}{2})\,\om^2\, x^2
  +  (\frac{1}{2})\,\frac{g}{x^2} \,,\quad g>0\,,
$$
representing an harmonic oscillator with a centripetal barrier, is
known as the isotonic oscillator \cite{WeJo79,Zhu87}. It is
important because is endowed with properties closely related with
those of the harmonic oscillator. At the classical level it leads
to periodic solutions with the same period (isochronous potential
\cite{ChaVe05}) and at the quantum level it is Schr\"odinger
solvable, the Hamiltonian is factorizable \cite{InHu51} and the
energy spectrum is equidistant. Moreover, it is related with
supersymmetric quantum mechanics \cite{Ca95}. The two-dimensional
version is superintegrable and corresponds to the so-called
Smorodinsky-Winternitz  system and the centripetal term relates it
with the Calogero-Moser system. In addition to all these
theoretical properties, this particular nonlinear oscillator is
important in quantum optics and in the theory of coherent states
\cite{WaLiZh00,ThSa04}.

The classical Lagrange equation, that is given by
$$
  \ddot{x} + \om^2\,x - \frac{g}{x^3} = 0 \,,
$$
is a particular case of the Pinney-Ermakov equation \cite{Pi50}.
It can be exactly solved and the solution is given by
$$
  x = \frac{1}{\om\,A}\,\sqrt{(\om^2A^4 - g)\,\sin^2(\om t + \phi) + g\,}\,.
$$
showing explicitly the above mentioned periodicity.  At the
quantum level it is convenient to write $g$ in the way $g=m(m+1)$
(the constant $g$ should be greater than $-1/4$ and $m$ can be any
real number but we will always take it as non-negative) so that
the Schr\"odinger equation takes the form
$$
  \frac{d^2}{dx^2}\,\Psi - \Bigl[\,\om^2\, x^2  +
  \frac{m(m+1)}{x^2}\,\Bigr] \Psi + 2 E\Psi = 0 \,.
$$
where we assume $\hbar=1$ for easy of notation.
The simplest solution, representing the ground state, is
$$
  \Psi_0 = N_0\, x^{1 + m}\, \exp(-\,\frac{1}{2}\om\,x^2)
  \,,\quad E_0 = \bigl(\frac{3}{2} + m\bigr)\,\om   \,.
$$
and the all the other wave functions and eigenenergies are given
by
$$
  \Psi_{2n} = N_{2n}\, x^{1 + m}\,P_{2n}(x)\,\exp(-\,\frac{1}{2}\om\,x^2)
  \,,\quad E_{2n} = E_0 + 2\,n\,\om \,,
$$
where $P_{2n}(x)$ is a polynomial of order $2n$ with only even
powers of $x$ and $N_{2n}$ denotes the normalization constant. The energy
spectrum is equidistant since
$$
  E_{2n+2} = E_{2n} + 2\,\om \,,\quad n=0,1,2,\dots
$$
Nevertheless, the height $\Delta E$ of the energy steps is twice
that of the simple harmonic oscillator $U_0$. In fact, it seems as
if half of the levels (those with an odd number of nodes) have
disappeared.

\section{A new solvable potential}

We now turn our attention to the study of the one-dimensional
quantum system described by the following potential
$$
  U_{0a}(x) = U_0(x) + U_a(x) = (\frac{1}{2})\,\Bigl[\,\om^2\, x^2
  + 2 g_a\,\frac{x^2 - a^2}{(x^2 + a^2)^2}\,\Bigr],\quad g_a>0\,,
$$
where $a$ is a positive real parameter. The reason for this
particular algebraic expression is that the new additional term
can be written as the sum of two centripetal barriers in the
complex plane
$$
  2\,\frac{x^2 - a^2}{(x^2 + a^2)^2} =
  \frac{1}{(x + \ii a)^2} + \frac{1}{(x - \ii a)^2} \,,
$$
so it is a rational potential with two imaginary poles symmetric
with respect the origin. Actually, if $g_a$ remains constant, then
when $a$ goes to zero and to $\infty$ the potential $U_{0a}(x)$
converges to the isotonic and the harmonic oscillator,
respectively.

The first important property is that if the coefficient $g_a$ is
not arbitrary but given by $g_a = 2\,\om\,a^2 (1 + 2\,\om\,a^2)$
then the Schr\"odinger equation, that takes the form
\begin{equation}
  \frac{d^2}{dx^2}\,\Psi - \Bigl[\,\om^2\,x^2
  + 4\,\om\,a^2  (1 + 2\,\om\, a^2)\,\frac{x^2 - a^2}{(x^2 + 
a^2)^2}\,\Bigr] \Psi
  + 2 E\Psi = 0  \,, \label{EqScha}
\end{equation}
admits the following solution
\begin{equation}
  \Psi_0 =  \frac{N_0}{(a^2 + x^2)^{2\,\om\,a^2}}\ \exp(-\,\frac{1}{2}\om\,x^2)
  \,,\quad E_0 = \frac{1}{2}\,\om - (2\,\om\,a)^2   \,.
\end{equation}
It represents the ground level ($\Psi_0$ has no nodes) and it
clearly shows that when $a\to 0$ then the corresponding wave
function and ground level energy of the linear oscillator is
obtained. We also note that the above expression  $g_a =
2\,\om\,a^2  (1 + 2\,\om\,a^2)$ can be alternatively written as
$g_a = m_a(m_a+1)$ with $m_a = 2\,\om\,a^2$.

  In the following we will focus our study on the particular case
$a^2=1/2$, $m_a=\om$ (see Figure 1) represented by the equation
\begin{equation}
  \frac{d^2}{dx^2}\,\Psi - \left[x^2 + 8\,\frac{2x^2-1}{(2x^2+1)^2}\right]\Psi
  + 2 E\Psi = 0   \label{EqScha12}
\end{equation}
so that $\Psi_0$ and $E_0$ become
\begin{equation}
  \Psi_0 (x)= \frac{N_0}{1 + 2 x^2}\ \exp(-\,\frac{1}{2}x^2)
  \,,\quad E_0 = - \,\frac{3}{2} \,,
  \label{Psi0E0}
\end{equation}
where we have assumed $\om=1$ for easy of notation.

As a method for obtaining all the other eigenstates we assume that
the functions $\Psi(x)$ can be factorized in the form
$$
  \Psi(x) = F(x)\,\Psi_0(x)
$$
and, as the lowest energy is $E_0=-3/2$, it seems appropriate to
write the general energy $E$ as follows
$$
  E= -\,\frac{3}{2} + e\,.
$$
Then the Schr\"odinger equation (\ref{EqScha12}) leads to the
following equation for the function  $F(x)$
\begin{equation}
  (1+2x^2)\,F'' - 2x\,(5+2x^2)\,F' + 2\,e\,(1+2x^2)\,F = 0\,.
\label{EqF}
\end{equation}

  Since the  origin $x=0$ is an ordinary point we expect an
analytic solution $F$ with a power series expansion convergent in
the interval $(-R,R)$  with the radius of convergence $R$ given by
$R=1/\sqrt{2}\,$
$$
  F(x)= \sum_{n=0}^{\infty} p_n x^n
   = p_0 + p_1 x + p_2 x^2 + p_3 x^3 + \dots
$$
that when replaced in (\ref{EqF}) leads to
\begin{eqnarray}
  &&2 (p_2 + e p_0) + [ 6 p_3 + 2 (e - 5) p_1] \,x
   + \sum_{\quad m=0}^{\quad\infty}\biggl[ \, (m+4)(m+3)\,p_{m+4}\cr
  &&+  2\bigl[\, (m+2)(m-4)  +  e\,\bigr]\, p_{m+2}
  + 4\bigl( e - m\,\bigr)\,p_m\,\biggr]\,x^{m+2} = 0 \,.
\end{eqnarray}
Therefore, we first obtain the following two relations for the
first coefficients $p_0,p_2$ and $p_1,p_3$
$$
  p_2 + e\,p_0 = 0 \,,\qquad  3 p_3 + (e - 5)\,p_1 = 0\,,
$$
and then the general recursion relation
$$
  (m+4)(m+3)\,p_{m+4} + 2 \bigl[\,(m+2)(m-4) + e\,\bigr]\, p_{m+2}
  +\,4(e -  m)\,p_m  = 0\,,\quad   m=0,1,2,\ldots\,.
$$
Hence the recurrence relation involve three different terms
($p_{m+4}$ depends of both $p_{m+2}$ and $p_m$) and is constrained
by the two first relations that are not included in the general
rule. In any case even coefficients are related among themselves
and the same is true for odd coefficients. The general solution
will be obtained by fixing the values of $p_0$ and $p_1$, i.e. the
value $F(0)$ and $F'(0)$. The solution determined by $F(0)=1$,
$F'(0)=0$ is even (only contains even powers) while the one
determined by $F(0)=0$, $F'(0)=1$ only contains odd powers of $x$.
Actually the expressions of the first coefficients, in terms of
$p_0$ and $p_1$, are given by
$$\begin{array}{rl}
  &p_2 = -\, e \,p_0  \cr
  &p_4 =  \frac{2^2}{4\,!}(e - 10) e \,p_0   \cr
  &p_6 = - \frac{2^3}{6\,!}(e - 26)(e - 4) e \,p_0
\end{array}\,\qquad
\begin{array}{rl}
  &p_8 =  \frac{2^4}{8\,!}(e - 50)(e - 6)(e - 4) e \,p_0   \cr
  &p_{10} = -\frac{2^5}{10\,!}(e - 82)(e - 8)(e - 6)(e - 4) e \,p_0   \cr
  & \dots\dots\dots \dots\dots\dots
  \end{array}
$$
and
$$\begin{array}{rl}
  &p_3 =   - \frac{2}{3\,!}(e-5)  \,p_1  \cr
  &p_5 = \frac{2^2}{5\,!}(e - 17) (e - 3) \,p_1  \cr
  &p_7 = - \frac{2^3}{7\,!} (e - 37)(e - 5) (e - 3) \,p_1
\end{array} 
\begin{array}{rl}
  &p_9 = \frac{2^4}{9\,!} (e - 65)(e - 7) (e - 5) (e - 3) \,p_1  \cr
  &p_{11} = -  \frac{2^5}{11\,!} (e - 101)(e-9)(e-7) (e-5) (e-3) \,p_1  \cr
  & \dots\dots\dots \dots\dots\dots
  \end{array}
$$
In the particular case in which  $e$ is an even integer number $e
= 2 k$, with $k = 3,4,5,\dots$, all the coefficients $p_{2(k+r)}$
vanish and the series reduces to a polynomial of degree $k$ in
powers of $x^2$, $P_{2k}(x)$. Similarly, when $e$ is an odd
integer number $e = 2 k+1$, with $k = 2,3,4,\dots$, all the
coefficients $p_{2(k+r)+1}$ with $r>1$ vanish, and  the series reduces
to a polynomial $P_{2k+1}$ with only odd powers of $x$. The first
polynomial solutions are:
$$\begin{array}{rl}
  P_0 &= 1    \cr
  P_4 &= 1 - 4 x^2 - 4 x^4   \cr
  P_6 &= 1 - 6 x^2 - 4 x^4 + \frac83 x^6 \cr
  P_8 &= 1 - 8 x^2 - \frac83 x^4 + \frac{32}5 x^6 - \frac{16}{15} x^8 \cr
  P_{10} &= 1 - 10 x^2 + \frac{32}3 x^6 - \frac{80}{21} x^8 + 
\frac{32}{105} x^{10}
\end{array} \quad
\begin{array}{rl}
  P_3 &= x + \frac23 x^3   \cr
  P_5 &= x - \frac45 x^5 \cr
  P_7 &= x - \frac23 x^3 - \frac43 x^5 + \frac8{21} x^7   \cr
  P_9 &= x - \frac43 x^3 - \frac85 x^5 + \frac{16}{15} x^7 - \frac{16}{135} x^9
\end{array}
$$
Two important properties are: First, there are no polynomial
solutions of degree $k=1$ and $k=2$ and, second, all of them have
two complex conjugate roots, so that $P_3$ has an unique real
zero, $P_4$ has two real zeros and, in the general case, $P_{2k}$
and $P_{2k+1}$ with $k>1$ have $2k-2$ and $2k-1$ real zeros
respectively.

Another remarkable property is that when the polynomials $P_n$ are
expressed as linear combination of Hermite polynomials then only a
finite number of terms are different from zero. We have obtained
the following relations for the first cases
$$\begin{array}{rl}
  P_4 &= - 4 H_0 - 4 H_2 - \frac14 H_4     \cr
  P_6 &=  3 H_2 + H_4 +\frac 1{24} H_6     \cr
  P_8 &= - \frac23 H_4 - \frac2{15} H_6 - \frac1{240} H_8     \cr
  P_{10} &=  \frac1{12} H_6 + \frac1{84} H_8 + \frac 1{3360} H_{10}
\end{array}
\qquad
\begin{array}{rl}
  P_3 &=  H_1 + \frac1{12} H_3      \cr
  P_5 &= - H_1 -  \frac12 H_3 - \frac1{40} H_5    \cr
  P_7 &= \frac13 H_3 + \frac1{12} H_5 + \frac1{336} H_7    \cr
  P_9 &= - \frac1{20} H_5 - \frac 1{120} H_7 - \frac 1{4320} H_9
\end{array}
$$
These particular relations clearly suggest that each polynomial
$P_{2k}$ can be written as a linear combination of $H_{2k}$ and
the two previous even Hermite polynomials, $H_{2k-2}$ and
$H_{2k-4}$; similarly the odd polynomial $P_{2k+1}$ appears as a
linear combination of only $H_{2k+1}$, $H_{2k-1}$ and $H_{2k-3}$.

\section{Wave functions and orthogonality relations}

The differential equation (\ref{EqF}) is not in selfadjoint form
because, if we denote by $a_0,a_1$ and $a_2$ the three
coefficients of the equation, we have
$$
   a_0 = 1+2x^2 \,,\quad a_1 = - 2x\,(5+2x^2)\,,\quad
    \frac{d a_0}{dx}\ne  a_1\,.
$$
However, use can be made of an integrating factor $\mu(x)$ such
that
$$
  \frac{d}{dx} [{\mu(x)a_0(x)}]  = \mu(x)a_1(x)\,.
$$
so that $\mu(x)$ is given by
$$
  \mu(x) = (\frac{1}{a_0})\,e^{{\int}(a_1/a_0)\,dx}
  = \frac{e^{-x^2}}{(1 + 2\,x^2)^3}\,,
$$
in such a way that (\ref{EqF}) becomes
$$
\frac{d}{dx}\Bigl[\,p(x)\,\frac{dF}{dx}\,\Bigr] + 2e\,r(x)\,F = 0 \,,
$$
where the two functions $p=p(x)$ and $r=r(x)$ are given by
$$
  p(x) = \frac{e^{-x^2}}{(1 + 2\,x^2)^2} \,,\qquad
  r(x) = \frac{e^{-x^2}}{(1 + 2\,x^2)^2} \,.
$$
This equation, with the appropriate conditions for the behaviour
of the solutions at the end points, constitutes a Sturm--Liouville
(S-L) problem, defined in the real line $\mathbb{R}$. According to
this, the eigenfunctions of the S-L problem are orthogonal with
respect to the weight function $r= e^{-x^2}/(1 + 2\,x^2)^2$, and,
in particular, the polynomial solutions $P_m$, $m=0,3,4,\dots$, of
the differential equation (\ref{EqF}), satisfy the orthogonality
conditions
$$
  \int_{-\infty}^{\infty} P_m(x)\,P_n(x)\,\frac{e^{-x^2}}
  {(1 + 2\,x^2)^2} \,\, dx = 0
  \,,\quad m\,\ne\,n \,.
$$

Let us now return to the Eq. (\ref{EqF}) and suppose for $F$ the
following factorization
$$
  F' = (1+2x^2)\,G  \,.
$$
Then we arrive after some calculus (we omit the details) to the
following equation for the function $G$
$$
  G'' - 2x\,G' + 2 (e-3)\,G = 0  \,,
$$
that means that the derivative $P_n'(x)$ of the polynomial
$P_n(x)$ must satisfy
$$
  P_n' = (1+2x^2)\,H_{n-3}\,,\quad e = n \,,\quad n=3,4,5,\dots
$$
(up to a multiplicative constant). At this point we recall the
following two properties of the Hermite polynomials
\begin{eqnarray}
  &(i)&\quad  2 x H_m = H_{m+1} + 2 m H_{m-1}  \cr
  &(ii)&\quad H_m'  = 2 m H_{m-1}  \nonumber
\end{eqnarray}
Then making use of (i) we arrive to
$$
  P_n' = \frac{1}{2}\,[H_{n-1} + 4 (n-2)\,H_{n-3} + 4
  (n-3)(n-4)\,H_{n-5}] \,,
$$
and making use of (ii) and then integrating we obtain
$$
  P_n = \frac{1}{4 n}\,\Bigl[ H_n + (4n)\,H_{n-2} + (4n)
  (n-3)\,H_{n-4}\Bigr] \,.
$$
It seems convenient to multiply $P_n$ by $4n$ and introduce the
new family of polynomials ${\cal P}_n$ defined in the form
\begin{equation}
  {\cal P}_n = H_n + 4\,n\, H_{n-2} + 4\,n\,(n-3)\, H_{n-4}
  \,,\quad n=3,4,5,\dots  \label{DefdePn}
\end{equation}
so that the coefficient of $H_n$ (the dominant term in the expression
of ${\cal P}_n$) reduces to unity.
\begin{proposicion}
The following equality holds
$$
  \frac{2n\,P_n\,e^{-x^2}}{(1+2x^2)^2} = -\,\frac{d}{dx}
  \Bigl[ \frac{H_{n-3}}{1+2x^2} \,e^{-x^2}\Bigr] \,,\quad n=3,4,5,\dots
$$
\end{proposicion}
{\it Proof:} This statement is proven just by making the calculus.

Now we can write
$$
  \int_{-\infty}^{\infty} \frac{{\cal P}_m(x)\,{\cal P}_n(x)}
   {(1 + 2\,x^2)^2} \,\,e^{-x^2} \,\, dx
  = -\,(16 n m) \int_{-\infty}^{\infty} \frac{1}{2m}\frac{d}{dx}
  \Bigl[ \frac{H_{m-3}}{1+2x^2} \,e^{-x^2}\Bigr]\,P_n(x) \,\, dx
$$
and integrating by parts we arrive to
\begin{eqnarray}
   \int_{-\infty}^{\infty} \frac{{\cal P}_m(x)\,{\cal P}_n(x)}
   {(1 + 2\,x^2)^2} \,\,e^{-x^2} \,\, dx
  &=& (8 n) \int_{-\infty}^{\infty} \Bigl[ \frac{H_{m-3}}{1+2x^2}
  \,e^{-x^2}\Bigr]\,P'_n(x) \,\,dx \cr
   &=& (8 n) \int_{-\infty}^{\infty}  H_{m-3}\,H_{n-3} \,e^{-x^2} \,\,dx \cr
  &=& \delta_{mn} \,(8 n) \,[\,2^{n-3}\,(n-3)\,!\,\sqrt{\pi} \,]
\end{eqnarray}
So, we relate the problem of normalization with the standard
problem of Hermite polynomials. Hence the orthogonality conditions
for the family ${\cal P}_n(x)$  read:
$$
  \int_{-\infty}^{\infty} {\cal P}_m(x)\,{\cal P}_n(x) \,r(x)\,dx
  = \delta_{mn}\int_{-\infty}^{\infty} \frac{[\,{\cal P}_n(x)\,]^2}
  {(1 + 2\,x^2)^2} \,\,e^{-x^2} \,\, dx
  = k_n\,\bigl(\,2^n\,n\,!\,\sqrt{\pi}\,\bigr) \,,
$$
where the proportionality constant $k_n$ is given by
$$ k_n = \frac{1}{(n-1)(n-2)}\,.
$$
See Figures 2a and 2b for the plot of some of these polynomials.

If we define the {\sl {\cal P}-Hermite functions \/} $\Psi_m$ by
$$
  \Psi_m(x) = \frac{{\cal P}_m(x)}{(1 + 2\,x^2)}\,\,e^{-(1/2)\,x^2}
  \,,\quad  m=0,3,4,\dots
$$
then the above property admits the following alternative form:
$$
  \int_{-\infty}^{\infty} \Psi_m(x)\,\Psi_n(x) \,\, dx = 0
  \,,\quad m\,\ne\,n \,.
$$

In summary, the  eigenfunctions corresponding to the lowest energy
levels are:
$$
\begin{array}{ll}
   \Psi_0(x) = N_0\,\dfrac{{\cal P}_0(x)}{(1 + 2\,x^2)}\,\,e^{-(1/2)\,x^2}
   \,,& E_0 = -\,3/2  \cr
   \Psi_3(x) = N_3\,\dfrac{{\cal P}_3(x)}{(1 + 2\,x^2)}\,\,e^{-(1/2)\,x^2}
   \,,& E_3 = 3/2 = -\,3/2 + 3   \cr
   \Psi_4(x) = N_4\,\dfrac{{\cal P}_4(x)}{(1 + 2\,x^2)}\,\,e^{-(1/2)\,x^2}
   \,,& E_4 = 5/2 = -\,3/2 + 4   \cr
   \Psi_5(x) = N_5\,\dfrac{{\cal P}_5(x)}{(1 + 2\,x^2)}\,\,e^{-(1/2)\,x^2}
   \,,& E_5 = 7/2= -\,3/2 + 5
\end{array}
$$
where the normalization constant is
$$
  N_k=\left[\frac{(k-1)(k-2)}{2^k\,k\,!\,\sqrt{\pi}}\right]^{1/2}\,.
$$
The first three wave functions, $\Psi_0(x)$, $\Psi_3(x)$,
$\Psi_4(x)$, together with the corresponding wave functions of the
harmonic oscillator $\Phi_0(x)$, $\Phi_1(x)$, $\Phi_2(x)$, are
plotted in Figures 3, 4 and 5.

The energy $E_0$ of the ground state $\Psi_0(x)$ has been singled
out of all the other values and moved into the smaller value $E_0
= -\,3/2$. The rest of the energy spectrum consists, as in the
pure harmonic case, of an infinite set of equidistant energy
levels
$$
  E_{n+1} = E_n + 1 \,,\quad n=3,4,5,\dots
$$

Let us close this section with two comments on the new family of
polynomials we have obtained. First the definition (\ref{DefdePn})
of ${\cal P}_n$ as a linear combination of {\sl only three}
Hermite polynomials can be considered as a particular case of a
situation known as a {\sl special linear combinations of
orthogonal polynomials} (see \cite{Sze75,Gr04} and references
therein). Finally, let us mention that taking into account the
``Rodrigues formula'' for the Hermite polynomials
$$
  H_n(x) = (-1)^n\,e^{x^2}\,\frac{d}{dx^n}\,e^{-x^2}
$$
we obtain
$$
  {\cal P}_n(x) = (-1)^n\,e^{x^2}\,\Bigl[\,\frac{d^n}{dx^n}
  + 4 n \,\frac{d^{n-2}}{dx^{n-2}}
  + 4\,n\,(n-3)\, \frac{d^{n-4}}{dx^{n-4}}\,\Bigr]\,e^{-x^2} \,,
$$
that must be considered as the ``Rodrigues formula'' for this new
family of orthogonal polynomials.

\section{Final comments and outlook}

We have proved that the potential $U_{0a}(x)$ can be exactly
solved in the particular case $a^2=1/2$ and also that it possesses
two very remarkable properties.  First, the fundamental level
$\Psi_0$ has an energy $E_0$ that is lower than in the pure
harmonic case and, in a sense, is isolated of all the other
values. Second, the rest of the energy spectrum is endowed with
the equidistance property. Concerning the general case, with an
arbitrary value for the parameter $a$, we have only obtained the
expression for the fundamental level ($\Psi_0(a),E_0(a)$). The
resolution of the general case remains as an open question that
deserves be studied. Finally let us also mention that the analysis
of this potential using the supersymmetric quantum mechanics as an
approach also seems an interesting matter to be studied.

\section*{Acknowledgments}
The paper was finished when one of authors (A.P.) was the guest of
Max-Planck-Institut f\"ur Mathematik (Bonn); he would like to
thank the staff of MPI for hospitality. We also thank M. Alfaro
for discussions on the properties of orthogonal polynomials.
Support of projects MTM-2006-10531, FPA-2003-02948, E23/1 (DGA),
MTM-2005-09183, and VA-013C05 is acknowledged.


\vfill\eject

\section*{Figure Captions}

\begin{itemize}

\item{}{\sc Figure 1}. Plot of the potential $U_{0a}(x)$ for
$\om=1$ as a function of $x$ for $a^2=1/2$ (continuous line)
together with the plot of the harmonic oscillator (dash line).
The main difference lies in the form of the minimum that is much
deeper in the $U_{0a}(x)$ case than in the linear case.
Nevertheless, for great values of $\left| x\right|$ the two
functions have rather the same form.

\item{}{\sc Figure 2}. (2a) Polynomials ${\cal P}_3$ (dash line)
and ${\cal P}_4$ (continuous line). $P_3$ has a unique real zero
at the origin and $P_4$ has two real zeros (symmetric with respect
the origin). (2b) Polynomials ${\cal P}_5$ (dash line) and ${\cal
P}_6$ (continuous line). $P_5$ has three real zeros  (the origin
and two other placed symmetric) and $P_6$ has four zeros (two
positive and two negative)..

\item{}{\sc Figure 3}. Wave function $\Psi_0$ (continuous line)
and wave function $\Phi_0$ of the harmonic oscillator (dash line).

\item{}{\sc Figure 4}. Wave function $\Psi_3$ (continuous line)
and wave function $\Phi_1$ of the harmonic oscillator (dash line).

\item{}{\sc Figure 5}. Wave function $\Psi_4$ (continuous line)
and wave function $\Phi_2$ of the harmonic oscillator (dash line).

\end{itemize}

\vfill\eject

\begin{figure}
\epsfbox{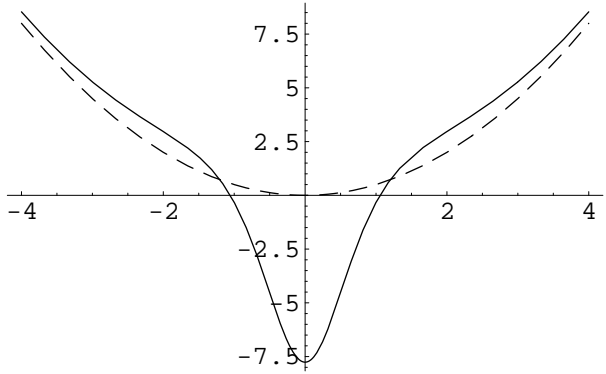}

\caption{Plot of the potential $U_{0a}(x)$ for
$\om=1$ as a function of $x$ for $a^2=1/2$ (continuous line)
together with the plot of the harmonic oscillator (dash line).
The main difference lies in the form of the minimum that is much
deeper in the $U_{0a}(x)$ case than in the linear case.
Nevertheless, for great values of $\left| x\right|$ the two
functions have rather the same form. }
\end{figure}

\smallskip
\begin{figure}
\epsfbox{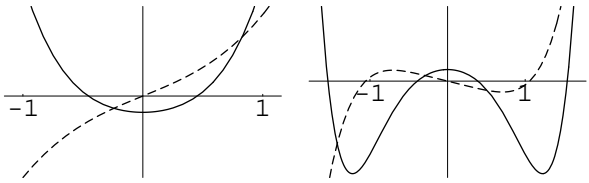}

\caption{(2a) Polynomials ${\cal P}_3$ (dash line) and ${\cal
P}_4$ (continuous line). $P_3$ has a unique real zero at the
origin and $P_4$ has two real zeros (symmetric with respect the
origin). (2b) Polynomials ${\cal P}_5$ (dash line) and ${\cal
P}_6$ (continuous line). $P_5$ has three real zeros  (the origin
and two other placed symmetric) and $P_6$ has four zeros (two
positive and two negative).}
\end{figure}

\begin{figure}
\epsfbox{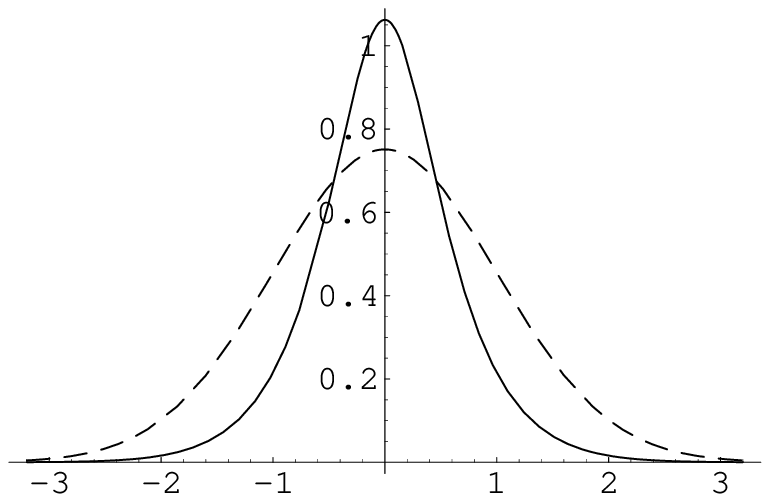}

\caption{ Wave function $\Psi_0$ (continuous line) and wave
function $\Phi_0$ of the harmonic oscillator (dash line).}
\end{figure}

\begin{figure}
\epsfbox{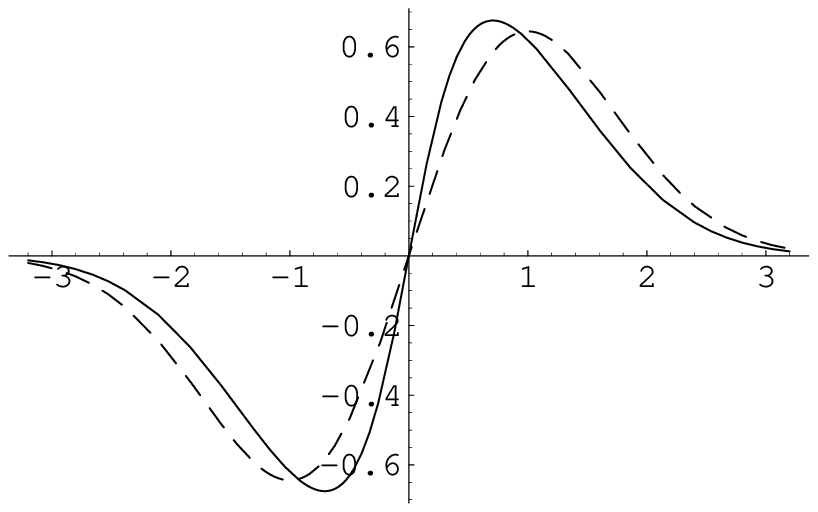}

\caption{ Wave function $\Psi_3$ (continuous line) and wave
function $\Phi_1$ of the harmonic oscillator (dash line).}
\end{figure}

\begin{figure}
\epsfbox{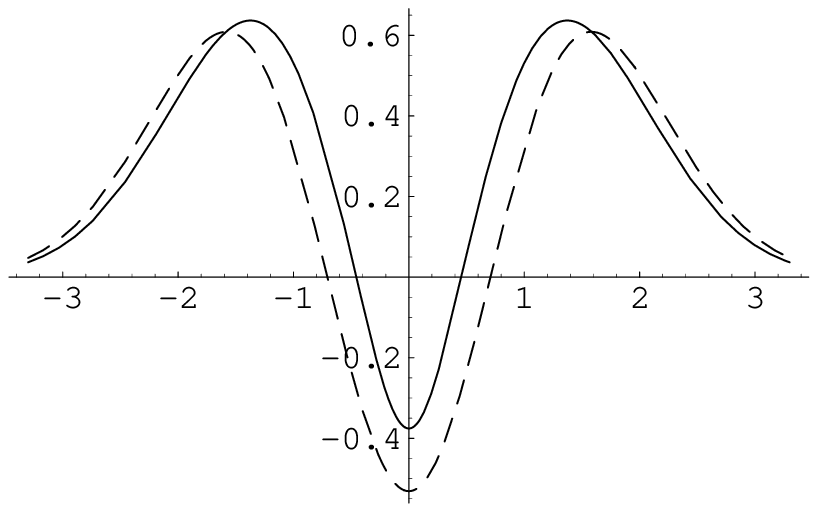}

\caption{ Wave function $\Psi_4$ (continuous line) and wave
function $\Phi_2$ of the harmonic oscillator (dash line).}
\end{figure}

\end{document}